\documentclass{aa}
\usepackage{graphicx}
\usepackage{verbatim}
\usepackage{flafter}
\usepackage{natbib}
\usepackage{journals}

\newcommand{\rod}{$\rho$ Oph D}
\begin{document}
 \title{3D Continuum radiative transfer in complex dust configurations 
        around young stellar objects and active nuclei}
 \subtitle{II. 3D Structure of the dense molecular cloud core \rod}
 \authorrunning{Steinacker et al}
 \titlerunning{3D Structure of $\rho$ Oph D}

 \author{
    J. Steinacker\inst{1}
    A. Bacmann\inst{2,3},
    Th. Henning\inst{1},
    R. Klessen\inst{4},
    M. Stickel\inst{1}
        }

 \offprints{stein@mpia.de}

 \institute{
    Max-Planck-Institut f\"ur Astronomie,
    K\"onigstuhl 17, D-69117 Heidelberg\\
    \email{stein@mpia.de}\\
    \email{henning@mpia.de}\\
    \email{stickel@mpia.de}
 \and
    European Southern Observatory,
    Karl-Schwarzschild-Str.2,
    D-85748 Garching
 \and
    Observatoire de Bordeaux,
    2 Rue de l'Observatoire, BP 89, 
    33270 Floirac\\
    \email{bacmann@obs.u-bordeaux1.fr}
 \and
    Astrophysikalisches Institut Potsdam,
    An der Sternwarte 16,
    14482 Potsdam\\
    \email{rklessen@aip.de}
            }

 \date{Received ; accepted }

 \abstract{
Constraints on the density and thermal 
3D structure of the dense molecular cloud core \rod\ are
derived from a detailed 3D radiative transfer modeling. 
Two ISOCAM images 
at 7 and 15 $\mu$m are fitted simultaneously
by representing the dust
distribution in the core with a series of 3D Gaussian
density profiles. 
Size, total
density, and position of the Gaussians are optimized by simulated annealing to
obtain a 2D column density map. The projected core density
has a complex
elongated pattern with two peaks.
We propose a new method to calculate an approximate
temperature in an externally illuminated complex 3D structure from a mean
optical depth.
This "$T_{\overline\tau}$"-method is applied to
a 1.3 mm map obtained with the IRAM 30m telescope to find the approximate 3D
density and temperature distribution of the core \rod.
The spatial 3D distribution deviates strongly
from spherical symmetry.
The elongated structure is in general agreement
with recent gravo-turbulent 
collapse calculations for molecular clouds.
We discuss possible ambiguities of the background determination
procedure, errors of the maps, the accuracy of the $T_{\overline\tau}$-method, and
the influence of the assumed dust particle sizes
and properties.
   }

   \maketitle
%
\section{Introduction}
Stars and planetary systems are believed to 
form within interstellar molecular clouds, from
the collapse of condensations of matter often called pre-stellar cores.
Despite recent observational and theoretical progress, the physical
processes leading to the formation of these cores, and
their collapse remain poorly known \citep[e.g.][]{2000prpl.conf...59A}. The study of 
pre-stellar cores is essential to the understanding of star formation, since
these objects represent the initial conditions of gravitational collapse
which influence the subsequent stages of the star-forming process. 

\subsection{Verification of 1D models of early star formation 
by observations of molecular cloud cores}
In the case of low-mass star-forming regions, it has traditionally
been thought that stars form from the inside-out collapse of singular
isothermal spheres that are initially in quasi-static equilibrium
\citep{1987ARA&A..25...23S},
supported against gravity by magnetic and thermal pressure and evolve
only due to slow ambipolar diffusion processes 
\citep{1991ApJ...373..169M}.
In this case, pre-stellar cores at the onset of collapse
are expected to follow a $1/r^2$ density profile at all radii $r$
\citep{1977ApJ...214..488S}.
Then they collapse dynamically
under their own weight and form a protostar, which subsequently accretes 
matter from  its surrounding envelope. 
 
In the last decade, much effort has been
put in the understanding of these cold ($T\sim$ 5-15 K) pre-stellar objects, 
which emit
most of their radiation in the (sub)millimeter domain with observations that 
benefitted from the development of millimeter bolometers,
sensitive line detector systems, and infrared cameras. Those
observations made it possible to derive line-of-sight integrated properties
of the
pre-stellar cores. Several studies using (sub-)millimeter continuum
\citep{1994MNRAS.268..276W,
       1996A&A...314..625A,
       1998A&A...336..150M,
       2004Laun} or
mid-infrared (MIR) absorption 
\citep{2000A&A...361..555B,
       1996A&A...315L.329A}
have in particular addressed the density structure.
Determining the density structure of pre-stellar objects is fundamental
to understand core and protostar evolution, since
it has been previously shown that the density
structure of {\it pre-stellar} objects influences the proto-stellar
accretion rate, i.e. after the formation of a central object. According to
\citet{1997A&A...323..549H} \citep[see also][]{1993ApJ...416..303F},
non-singular density profiles in pre-stellar cores
lead to accretion rates declining with time  in the proto-stellar stage,
whereas the accretion rate is expected to be constant if the core has
singular $r^{-2}$ profiles \citep{1977ApJ...214..488S, 1987ARA&A..25...23S}.

It was found from the mm and MIR maps that column density
profiles of pre-stellar objects are flattened out in their centers.
Interpreted by 1D singular isothermal
sphere models, the derived density distributions do not
follow the $r^{-2}$-profile from the \citet{1987ARA&A..25...23S}
model, though the profiles do have an $r^{-2}$ part. 

The fact that
some cores even present sharp edges at large radii 
\citep[$\sim 0.1$\,pc, see][]{2000A&A...361..555B} tends to show that the reservoir of mass available for 
subsequent accretion is limited. Therefore to some extent, the density 
structure of pre-stellar cores must govern stellar masses and influence
the initial mass function \citep{2003IAUS..221E.154U}. 

Using the 1D approximation,
several types of hydrostatic and quasi-static models have 
been proposed to
account for core formation and evolution.
Comparison of the density
profiles derived from the observations with the models has been used in
order to discriminate between various models
\citep[e.g.][]{1994MNRAS.268..276W,
       1998Obs...118..346W,
       1996A&A...314..625A,
       2000A&A...361..555B}.       
In most of these studies, circular (or elliptical) averages of the
dense core emission (either in the mm or in the MIR) have been
performed according to the apparent (projected) geometry of the object.
The azimuthal average was also used to
increase the signal-to-noise ratio. The profiles thus derived were
compared to various power law density models. It was found that the
ambipolar diffusion models of magnetically-supported cores provide the
best fits, requiring a background magnetic field ranging from 30 to
100 $\mu$G \citep{2003ApJ...592..233W}.

\subsection{Limitations of 1D models of early star formation}
Both from observational and from the theoretical side, 
there are many indications
that star formation is an intrinsically three-dimensional process.
The profile analysis was often performed in 1D or 2D because of the difficulty 
of obtaining information along the 3rd dimension, as well as the substantially 
enhanced numerical effort in 3D star-formation simulations and 3D radiative
transfer modeling. 

However, the increased sensitivity of
bolometer arrays as well as the observations of cores in absorption in the
mid-infrared 
\citep[a technique which is more sensitive to small column densities -
e.g.][]{2000A&A...361..555B, Stein04} 
reveals filamentary structures for the cores at large
scales, which significantly differ from near-spherical approximations
\citep{2004PhDT.........1A}.
Statistical analyses of core projected shapes
\citep{2001ApJ...551..387J, 2002ApJ...569..280J}
 also tend to show that cores are triaxial.
If 1D interpretations represent
in simple cases (i.e. for isolated cores) 
a good approximation of the cores, they can also lead to
erroneous density structures, as illustrated by \citet{2003ApJ...592..188B}.
More specifically, \citet{Stein04} used non-symmetric, filamentary
density distributions from a smooth particle hydrodynamics simulation
of a cloud core evolution to produce MIR and mm maps. By interpreting the
absorption maxima in MIR maps and emission maxima mm maps with a 1D model, 
they found density profiles in ostensible
agreement with standard 1D core formation
models, but 
in contradiction to the true underlying non-symmetric distribution. 

Moreover, the increased sensitivity and resolution
of upcoming telescopes will make it possible to see small-scale structures
in these objects and make more sophisticated models more and more necessary.
The 3D density structure and eventually its evolution in time is also of 
crucial interest for chemical models \citep[e.g.][]{2003ApJ...593..906A}.

\subsection{Gravo-turbulent star formation}
The hypothesis that star formation mostly occurs in magnetically
subcritical cores and is regulated by slow ambipolar diffusion processes
\citep{1987IAUS..122....7S}
has been challenged from
several sides \citep[see, e.g., the reviews by][]
{Larsen2003,
2004RvMP...76..125M}.
It is only applicable to isolated, single stars, while it
is known that the majority of stars form in small aggregates or large
clusters 
\citep{2001ApJ...553..744A,
       2003ARA&A..41...57L}.
Furthermore, there is both observational evidence
\citep{2004mim..proc..123C,
       2000prpl.conf...59A,
       2001ApJ...554..916B,
       2001ApJ...561..871H,
       2003ApJ...592..233W} and theoretical reasoning 
\citep[e.g.][]{1998ApJ...494..587N}
that cloud cores do not have
magnetic fields strong enough to support the core against gravitational
collapse.  Also, the long lifetimes implied by the quasi-static
phase of evolution in the model are currently discussed regarding, e.g.,
observational statistics of cloud cores 
\citep{1996A&A...313..269T,
       1999ApJS..123..233L,
       2002AJ....124.2756V} and chemical age considerations 
\citep{1998ARA&A..36..317V, 2000prpl.conf...29L}. 
\citet{TasMou04}, however, have argued that the observational data 
are consistent with the predictions of the theory of self-initiated, 
ambipolar-diffusion--controlled star formation.

An alternative to mediation by magnetic fields as controlling agent for stellar birth
is supersonic interstellar turbulence \citep{2004RvMP...76..125M}.  
The presence of highly
supersonic (and usually super-Alfv{\'e}nic) turbulent motions in
Galactic gas clouds is inferred from molecular line measurements.  The
observed line widths typically exceed the values for thermal line
broadening by far \citep{1993prpl.conf..125B}. Only on very
small scales and for the so-called quiescent cores does the total
linewidth become comparable to the thermal width alone. It is
important to note that turbulence usually carries sufficient energy
to counterbalance gravity on global scales. The presence of supersonic
turbulence therefore prevents contraction of the cloud as a whole. On
small scales, however, it may actually provoke localized collapse
\citep{1982ApJ...256..505H,
       1993ApJ...419L..29E,
       1995MNRAS.277..377P,
       1999ApJ...527..285B,
       2000ApJ...535..887K,
       1999ApJ...526..279P,
       2002ApJ...576..870P,
       2001ApJ...547..280H}.
Supersonic turbulence establishes a complex
network of interacting shocks, where converging flows generate regions
of enhanced density.  The compression can be sufficient for local
gravitational instability to set in.  The same random flow that
creates density enhancements in the first place, however, can also
disperse them again. For star formation to occur, the localized
compression must be fast enough for the contraction region to
`decouple' from the flow.

The theory of gravo-turbulent star formation predicts that molecular cloud
cores form at the stagnation points of the complex turbulent flow
pattern. Their density profiles are constant in the inner region,
exhibit an approximate power-law fall-off further out, and usually
appear truncated at some maximum radius
\citep[see][]{2003ApJ...592..188B}. Such configurations 
closely resemble equilibrium Bonnor-Ebert spheres 
\citep{1955ZA.....37..217E,
       1956MNRAS.116..351B},
which are popular models for interpreting observed cloud cores.
The best studied example probably is the Bok globule B68 
\citep{2001Natur.409..159A,
       2003ApJ...586..286L}.
However, supersonic turbulence does not
create hydrostatic equilibrium configurations.  Instead, the density
structure is transient and dynamically evolving, as the different
contributions to virial equilibrium do not balance
\citep{2003ApJ...585L.131V}. 
Numerical models predict that gravo-turbulent fragmentation
of molecular cloud material will lead to cores that are generally
triaxial and in many cases highly distorted.  Depending on the
projection angle, they often appear extremely elongated, being part of
a filamentary structure that may connect several objects \citep{Klessen04}. 

It is the goal of this paper to extend our understanding of pre-stellar
cores by entering the complex and difficult world of 3D modeling.
The target of our investigation is the pre-stellar core \rod\ located in the
$\rho$ Ophiuchi star-forming cluster.
This core was selected because it has been observed 
both in the MIR and mm range and the maps indicate a cloud structure
deviating from spherical symmetry.
We start by outlining the general strategy of the analysis in Sect.~2.
In Sect.~3,
we present results of a simultaneous fit of
ISOCAM images at 7 and 15 $\mu$m 
by a series of 2D Gaussian profiles
to obtain the number column density at each image pixel. 
The necessary background determination scheme is described and
discussed.
We compare the derived dust density to an IRAM 30m 1.3 mm map 
in Sect.~4.
Presenting and applying a new method to obtain 3D density and temperature
information of cores, we derive constraints on the structure of the
core \rod.
A discussion of ambiguities and errors of the obtained distributions
is given in Sect.~5.
In Sect.~6, we compare the derived spatial
distribution with distributions
typically obtained in gravo-turbulent simulations and conclude our
findings in Sect.~7.

\section{General strategy of the 3D analysis}
As the 3D analysis of molecular cloud cores based on continuum images
is new and includes several steps and approximations, we present
in this section a simplified overview of the applied method.
Details, tests, and further discussion of single points have been 
moved to the later sections for clarity.

First, we determine
the column density of the core
region for each pixel of the 7 and 15 $\mu$m images. In order to estimate the
background radiation that is absorbed by the core, we consider the
un-absorbed background radiation near the core and interpolate it
behind the core. The optical
depth is calculated for each pixel and, under the assumption of dust
particle size and opacities, converted to a dust column density.
To model the 3D distribution of the core, we use a series of
3D Gaussian distributions with axes aligned to the Cartesian axes.
The position, the half width at half maximum (HWHM) along the three axis, and the
density weight contain 7 free parameters for each Gaussian.
To model the column density, the position in the plane of sky, two 
HWHM, and the weight are varied to fit the images. The corresponding
$\chi^2$-function is minimized using simulated annealing.

The dust density and temperature 
distribution along the line of sight is determined by considering
mm emission of the dust particles absorbing in the MIR.
The emission seen in the 1.3 mm image is obtained from 
a line-of-sight integral
over the emission at each location in the core depending on both
the local dust density and temperature. By varying the position
and HWHM of the Gaussians along the line of sight, we can fit the
1.3 mm image to obtain the 3D density and temperature structure
applying again the simulated annealing optimization algorithm.

The calculation of the dust temperature distribution for a given
density distribution requires the application of a 3D continuum radiative
transfer (CRT) code.
As these computations are far too time-consuming to be repeated
many times, we propose and apply an
approximative relation to calculate the temperature $T$ at a location in
the core from a mean optical depth $\overline\tau$ at 7 $\mu$m obtained by a weighted
averaging over all directions. This $T_{\overline\tau}$-relation is obtained from
a few 3D CRT runs covering average and limiting density distributions.

\section{Determination of the dust number column density 
map from ISOCAM maps at 7 and 15 $\mu$m}
\subsection{Background determination}
To obtain a detailed dust column density map of the core, we use
the ISOCAM maps from 
\citet{1996A&A...315L.329A}, which also have been discussed in
\citet{1998sfis.conf..220A}.
The maps were
observed at a central wavelength of $\sim 7$\,$\mu$m (LW2 filter
[5--8.5\,$\mu$m]) and 15\,$\mu$m (LW3 filter [12--18\,$\mu$m]) with a
resolution of $6^{\prime \prime}$. 
The MIR emission seen in the $\rho$ Oph cloud is thought to arise from
very small grains or Polycyclic Aromatic Hydrocarbons (hereafter
PAHs) excited on the outside of the cloud by far-UV radiation. The PAHs
then re-emit the absorbed energy in the form of spectral bands in the MIR,
of which several are encompassed in the ISOCAM filters. Since the
extinction in the far-UV is very high and does not penetrate further than
$A_V\sim 1$ mag, the core itself should be free of PAH emission, the bulk of the
observed MIR emission arising from the outer parts of the cloud. For the
same reason, emission of the cold dust particles within the core in
the 7-15\,$\mu$m domain is highly unlikely (the core temperature is lower
than 20\,K). The radiation received can be modeled as follows (see also
Fig.~\ref{Sketch}):
\begin{figure}[t]
\vspace{8.5cm}
\includegraphics{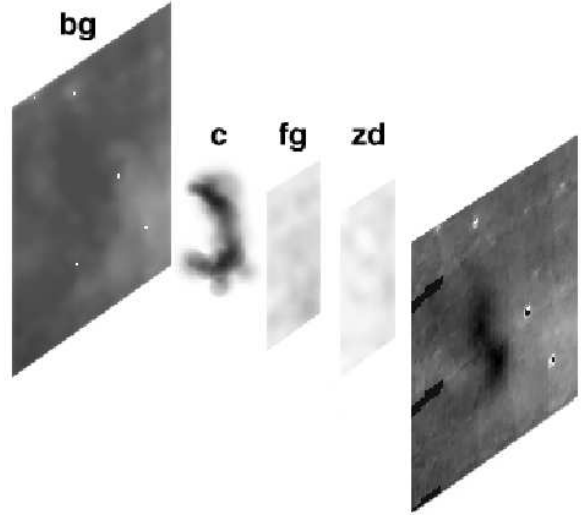}
\caption{Sketch of the possible layers of matter altering the
background radiation on its way through the dense core.
\label{Sketch}
}
\end{figure}
1. A background layer at the origin of excited PAH emission from the outer
cloud parts (labeled $bg$),
2. The dense core absorbing this background radiation in the MIR (labeled
$c$),
3. A foreground layer similar to the background layer containing absorbing
cold dust and
a layer of emitting small particles at the border of the cloud ($fg$),
and
4. A
layer of zodiacal dust with emission and absorption ($zd$).
With this approximative decomposition, the total intensity along the line
of sight can be written as
\begin{equation} \label{decomp}
 I_{obs}=\left(
 I_{bg}\ e^{-\tau_c}\ e^{-\tau_{fg}} + I_{fg}+I_{zd}
\right)\ e^{-\tau_{zd}}
\end{equation}
where $\tau$ is the optical depth.

The $\rho$ Oph cloud is illuminated from behind by the B2V-star HD147889
\citep{1999A&A...344..342L}
exciting the PAHs of the background layer. As there is no star in the
vicinity of the cloud illuminating the foreground layers, the PAH emission
arising from the foreground layer is supposed to be negligible.  This is
in agreement with the results of Bacmann et al. (2000) who found that the
entire foreground emission came from zodiacal dust. The zodiacal dust
emission  was determined
by \citet{1996A&A...315L.329A} to be 5.7 and 39 MJy/sr for the LW2
(7\,$\mu$m) and LW3 (15\,$\mu$m)
maps, respectively, and has been subtracted from the map.
Currently, a possible absorption by cloud dust in front of the core can
not be
distinguished unambiguously from the core absorption. We note that a small
part of the dust distribution derived from the maps might belong
to the foreground cloud rather than to the core itself. In that sense,
the derived dust densities will be an upper limit to the real densities.
                                                                                
To determine $\tau_c$ and the dust column density map, the background
behind
the core has to be interpolated from the observed background in the
vicinity of the core.
We assume that the background behind the core follows the gradients that
are
visible in the maps.
After excluding bad pixels containing contamination from previous
observations
and periodic remnants, we determine a first background approximation
by manually
masking the core area and interpolating the outer background into the
masked
area. The background at a given pixel was interpolated as the average of
all non-masked pixels within a given radius. This radius was chosen to be
1/3 of the map extent. This reproduces well the
background gradient throughout the map without emphasizing local noise
features.
After having defined this first-order background, we redefine the core
area
to be the range of pixels where the absorption of this background is
larger
than $1/e$.
Again, the nearby background is interpolated into this new core area.
We have verified the stability of the results of our fitting procedure by
varying
the averaging radius and the size and shape of the initial
manually-determined mask of the core area.

\subsection{Determination of the optical depth and column density map} 
The interpretation of the obtained optical depth along lines of sight
through the core is ambiguous.
The absorption coefficient of dust particles $\kappa$ is a function
of their chemical composition $C$, 
of their size $a$, the temperature $T$, the wavelength $\lambda$,
and (via $C$, $a$ and $T$) the location in space.
As the temperature range of roughly 5-20 K is small 
\citep[see, e.g.,][]{2004A&A...420.1009S}, 
we neglect temperature 
dependencies for the absorption coefficient.
As far as dust size is concerned, the dust may grow in the innermost parts
of the core changing both its size and chemical composition and
hence its optical properties.
A mixing of the dust particles due to gas motion
countering spatial dependencies
is possible but hard to model due to the complex turbulent motions
in these cores.
A minimum modeling of the dust evolution and the gas motion would
be required \citep[see][]{1994A&A...291..943O}, 
but it would introduce a wealth of more parameters into the fitting
procedure. 
Since in this paper, we would like to emphasize the 
detailed structure that was often neglected in former 
models,
we assume that absorption is dominated by dust of a typical
size $a$ and refer to a forth-coming paper dealing with
a possible spatial variation of the dust size and absorption
coefficient.

The dust number column density 
is calculated by ray-tracing the background intensity
through a given spatial 3D dust number
density distribution representing the core.
There are several useful basis functions that can be used to describe 
the spatial 3D dust
density distribution
$n(x,y,z)$ of the core in a series expansion in Cartesian coordinates.
Here, we choose a series of $g$ 3D Gaussians with their main axis
aligned with the coordinate axis, having the form
\begin{equation} \label{Gaussian}
  n(x,y,z) =
      \sum\limits_{i=1}^g
       G_i \exp
        \left[ 
          -d_i(x,y,z)
        \right]
\end{equation}
with the argument 
\begin{equation} \label{Argu}
d_i(x,y,z)=
\left(\frac{x-x_i}{a_i}\right)^2+
\left(\frac{y-y_i}{b_i}\right)^2+
\left(\frac{z-z_i}{c_i}\right)^2.
\end{equation}
The parameterization
contains $7*g$ free parameters (weights $G_i$, translations 
$x_i$, $y_i$, $z_i$, and size parameters 
$a_i$, $b_i$, $c_i$).
This representation has the advantage that 
the Gaussian profiles
are a good approximation for the flattened innermost part of
the cores that is seen for many cores \citep[for profiles 
see, e.g.,][]{2000A&A...361..555B}, 
and include the spherically symmetric case.
For the outer parts, the gradients of the density profile
have a large uncertainty as projection effects severly change
the profiles that are obtained from column density modeling
\citep[see, e.g.,][]{Stein04}. Therefore, it is currently
uncertain if the density follows a powerlaw or an exponential.
To keep the inversion method proposed 
in this paper numerically feasible, we have chosen
the number of Gaussians $g$ to be 30, and the implications of this
choice are discussed in Sect. 4.2 and Sect. 5.1.
We also have tested a series of Gaussians with axes inclined to the
line of sight and got the same overall column density
distribution with a lower number of Gaussians. 
The numerical effort is higher, however, since the formula for Gaussians
with axes inclined to the line of sight is substantially more complicated.
Moreover, for 
a given number density $n(x,y,z)$, conveniently, the number column density
\begin{eqnarray} \label{CD}
  N(y,z)&=&
      \int\limits_{-\infty}^{+\infty}
      dx\ n(x,y,z)\cr
        &=&
      \sqrt{\pi}
      \sum\limits_{i=1}^g
      a_iG_i
      \exp\left[
     -d(x_i,y,z)
      \right]
\end{eqnarray}
can be calculated 
analytically if the line of sight is chosen to be, e.g., along the x-axis.
An interclump medium represented by a constant density component has not
been added to the density as the Gaussians include this case for large
size parameters.
Assuming dust particles with an internal density of $\rho_d$, the 
corresponding total
dust mass in the core is
\begin{equation} \label{M}
M=\frac{4}{3}\pi^{5/2}a^3\rho_d
\sum_i^g G_ia_ib_ic_i.
\end{equation}
The optical depth between two points $x_1$ and $x_2$ on the line
of sight is
\begin{equation} \label{OD}
  \tau_x(x_1,x_2,y,z,\lambda)= 
      \sigma(\lambda)
      \int\limits_{x_1}^{x_2}
      dx\ n(x,y,z)
\end{equation}
with the absorption cross section
$\sigma(\lambda)$.
For a ray crossing the entire core, we find
\begin{equation} \label{OD1}
  \tau_x(-\infty,\infty,y,z,\lambda)=
      \sigma(\lambda) N(y,z).
\end{equation}
Summing over all pixels k and maps l, 
we define the $\chi^2$-function of the fit to be 
\begin{equation} \label{Chi2}
  \chi^2=
      \sum\limits_{l=1}^{m}
      \sum\limits_{k=1}^{N_l}
      \left(
         F_l(k)-D_l(k)
      \right)^2
\end{equation}
where $D_l$ contains the flux values of each pixel of the two images,
$F_l$ is the background flux after undergoing absorption due to
the model core, $m=2$ is the number of maps, and $N_l$ is the number of
pixels of map $l$.

The optimization of the 150 parameter dust number density by
fitting about 1430 pixels of the MIR maps was performed
with a code based on the simulated annealing algorithm 
\citep{1983Sci...220..671K,
       Vanderbilt,
       1994A&A...287..493T, 
       1996jqsrt..56..97S, 
       2003ApJ...583L..35S} which is able 
to both find minima in the
manifold parameter space and to find the deepest minima.
\begin{figure}[t]
\vspace{9.8cm}
\includegraphics{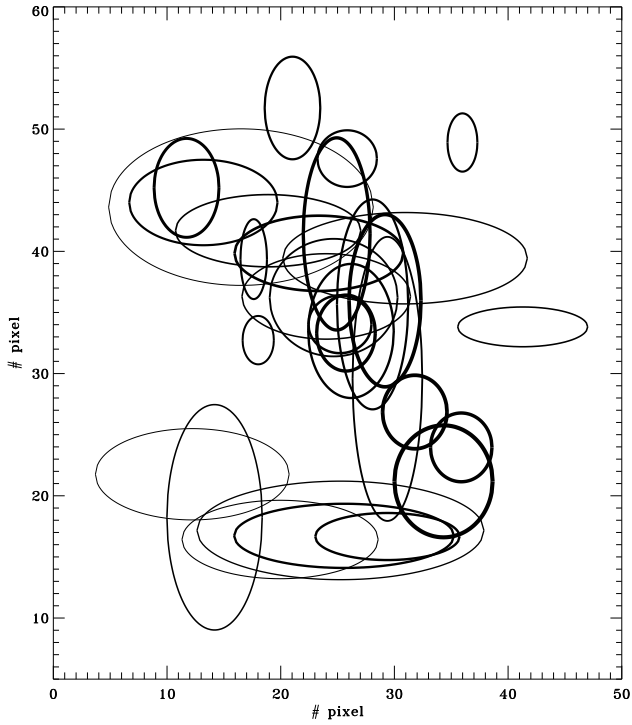}
\caption{Relative locations and sizes of the individual
Gaussians for the optimized number column density distribution. 
The ellipses are cuts of the Gaussians for $x=x_i$ and 
$n_i=G_i/e^2$.
The thick-lined ellipses indicate the most dense Gaussians.
\label{Gaussians}
}
\end{figure}
In Fig.~\ref{Gaussians}, 
the relative location and size of the individual 
Gaussians is indicated by
ellipses having HWHM of size $a_i$ and $b_i$.
The line thickness of the ellipses is logarithmically proportional
to the density weight of the Gaussians.
The general
form of the core
is an elongated structure with two peaks and a ratio of projected 
axis of about 4:1. The larger southern
peak is well described by three almost spherical Gaussians.
The northern peak has a complex shape and a correct representation
of its density structure
may require more Gaussians than used in this analysis.
\begin{figure}[t]
\vspace{9cm}
\includegraphics{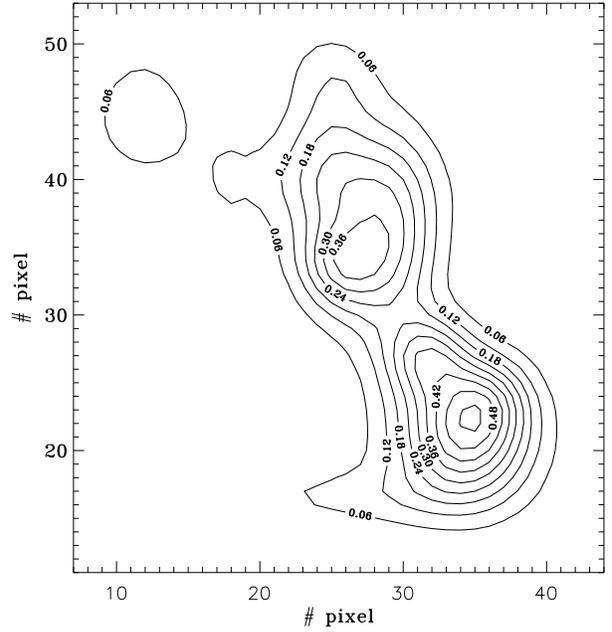}
\caption{Map of the optical depth $\tau_x$ at 7 $\mu$m
of the cloud core derived from
simultaneously fitting
ISOCAM maps at wavelengths of 7 and 15 $\mu$m.
Each pixel has a size of 960 AU.
\label{ndensmap}
}
\end{figure}
\begin{figure}[t]
\vspace{9cm}
\includegraphics{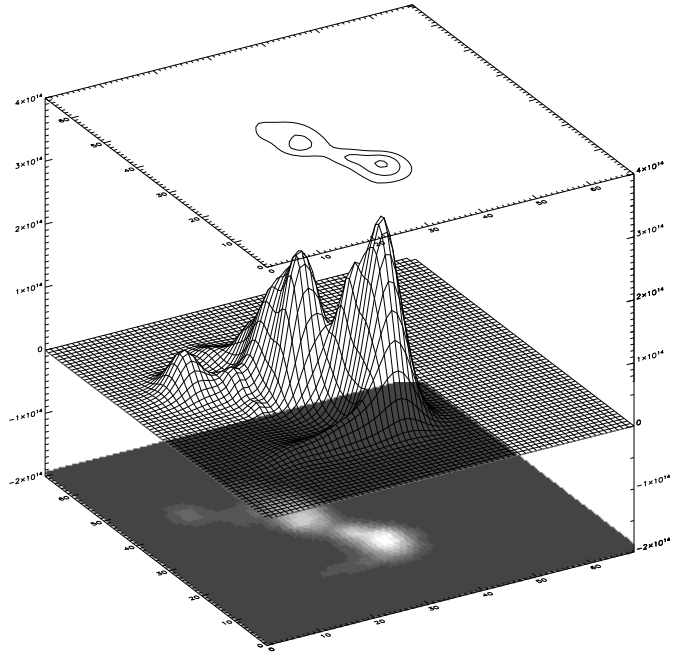}
\caption{Representations of the number column density
of the cloud core derived from
simultaneously fitting
ISOCAM maps at wavelengths of 7 and 15 $\mu$m.
\label{taumap}
}
\end{figure}

The number column density map is shown in Fig.~\ref{ndensmap} as a 
function of the pixel
index of the map (one pixel has a size of 960 AU).
The double-peaked structure is pronounced,
with the southern peak being the denser part.

In Fig.~\ref{taumap}, the optical depth map for 7 $\mu$m is shown
as a function of the pixel number. The contours are labeled with
the optical depth value $\tau_x$ along rays completely crossing the core,
reaching a maximum of about 0.5 in the southern
maximum and less than 0.4 in the northern maximum. The optical
depth does not exceed 0.14 in the region between the two maxima.

\begin{figure}[t]
\vspace{9cm}
\includegraphics{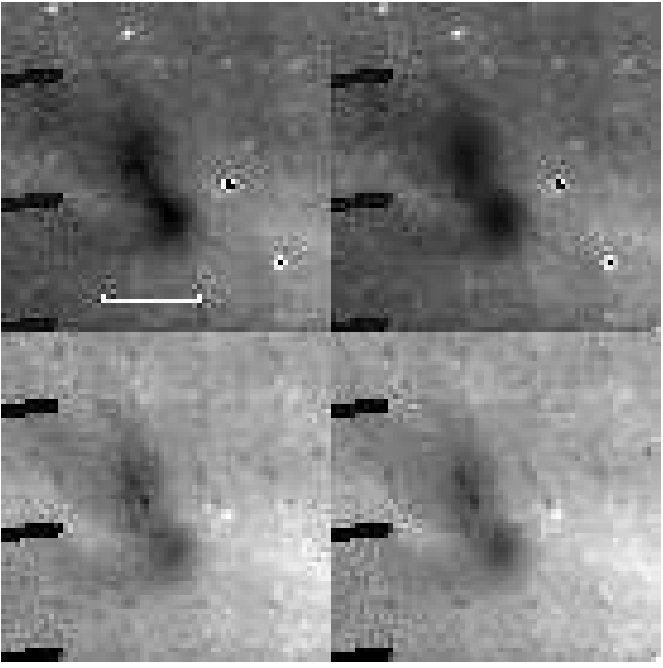}
\caption{ISOCAM images of the dense molecular cloud core \rod\ at 
$\lambda=7\ \mu$m (upper left) and $\lambda=15\ \mu$m (lower left) and maps resulting
from fitting all maps with the same number density distribution (right).
The white line in the upper left panel indicates a spatial length scale of
about 0.13 arcmin or 0.1 pc (20000 AU).
\label{7and15}
}
\end{figure}
The resulting fit maps using the optimized number density distribution are given
in Fig.~\ref{7and15} along with two of the original ISOCAM images
(upper and lower left panel shows 7 and 15 $\mu$m, respectively).
The white line
in the upper left panel represents a spatial scale of about 0.13 arcmin
or 0.1 pc (20000 AU).
The fit is not able to reproduce the small-scale noise that is
part of the real maps but resembles well the overall appearance of the
cloud core.

With 30 Gaussian basis functions involved, the fit is ambiguous.
The most obvious ambiguity when running the optimization procedure arises
from identical density structures where Gaussians just have been exchanged.
Moreover, there are several possibilities to arrange
the Gaussians to represent the complex extended double-peak structure
of such a core. These correspond to different local minima in the parameter
space and it requires a special minimization scheme like the
simulated annealing algorithm as used in this modeling 
to find the global minimum
of the $\chi^2$-function.
A more detailed error analysis is given in Sect.~5.

\section{Determination of constraints on the 3D density and
temperature distribution}
\subsection{The $T_{\overline\tau}$-method}
In this section, we want to use a mm map to 
further constrain the density and temperature 
distribution of the dust.
However,
the dust particles absorbing the radiation at 7 and 15 $\mu$m
are not obviously the ones dominating emission in the mm range.
We assume in first approximation that the absorption and emission
is dominated by particles of size a.
The absorption in the MIR is described by the 
optical depth along the line of sight
\begin{equation}
\tau(y,z)=
\sigma
\int\limits_{-\infty}^{+\infty}
n(x,y,z)\ dx =
\frac{\sigma\ M_c(y,z)}{m_d}=
\frac{3}{4\rho_d}\frac{Q}{a}M_c(y,z)
\end{equation}
with the total column dust mass $M_c$, the dust particle mass $m_d$,
and the absorption factor $Q$.
For pre-stellar cores, the dust sizes are presumably small enough so that
the Rayleigh limit $2\pi a \ll \lambda$ is fulfilled for wavelengths around
1 mm and commonly assumed chemical composition of astrophysical dust.
Hence, $Q\propto a$, and the optical depth is independent of the
dust size as long as the total dust mass is not changed.

The intensity of the mm emission is
\begin{equation}
I_{mm}(y,z)\propto
\sigma
\int\limits_{-\infty}^{+\infty}
n(x,y,z)\ B_\lambda[T(x,y,z)]\ dx
\end{equation}
where $B$ is the Planck function and $T$ is the temperature of the dust particles
(a detailed description of the mm emission
is given in Sect.~4.2).
The temperature of the dust particles can vary with particle size when the dust is
exposed to strong UV and optical radiation.
Pre-stellar cores like \rod\ are shielded 
from the short-wavelength
part of the spectrum which reduces differences in the dust temperature of
different dust sizes. For this application, we assume
a temperature depending on the location but not on grain size.
Therefore, the emission in the mm is also proportional to the optical depth and
it is reasonable to use the spatial dust particle
distribution found by fitting maps in the MIR to model
the mm emission.

To analyze the mm map, an important parameter is
the temperature of the dust 
in the core with a column density determined from the MIR fit.
Temperature distributions of embedded and non-embedded spherically symmetric
cores have been calculated by \citet{2003A&A...407..941S} ranging from 6.5
to 18 K. \citet{2001A&A...376..650Z} have
discussed both 1D temperature distributions and
temperatures in disk-like structures. 

The core \rod\ clearly has a more 
complex structure. 
The temperature of the dust at a certain location $\vec x$ within the core is 
determined by the local power density balance of incoming and outgoing radiation
\begin{eqnarray}\label{balance}
&&
\int\limits_0^\infty d\lambda\ 4\pi\ a^2\ Q_{abs}(\lambda)
\ n(\vec x)
\ B_\lambda[T(\vec x)] =\cr
&&
\int\limits_0^\infty d\lambda\ \pi\ a^2\ Q_{abs}(\lambda)
\ n(\vec x)\ 
\frac{1}{4\pi}
\int\limits_0^{2\pi} d\phi \int\limits_0^{\pi} d\theta \sin\theta\ 
I(\lambda,\vec x,\theta,\phi)
\end{eqnarray}
with the intensity $I$ and the direction described in spherical coordinates
$\theta$ and $\phi$.
Generally, the radiation 
intensity within the core depends both on the external irradiation and
the re-emission of the dust particles, coupling $I$ and $T$ over all
wavelengths, locations, and directions.
For a complex density structure like the core \rod,
a 3D CRT calculation
with self-consistent temperature iteration
is required. However,
the numerical effort of these codes strongly prohibits any fitting
of multi-parameter distributions, as the temperature distribution
has to be determined many thousand times in order to find the optimal
multi-parameter set in the solution space.

To estimate the range of the temperatures within \rod,
we calculated the radiation field emerging from several density
distributions with Gaussians being stretched and translated
along the line of sight, but with a fixed extent and column density
in the plane of
sky to fit the MIR map. The calculations have been performed by a
3D CRT code presented in \citet{2003A&A...401..405S} 
                         \citep[see also][]{2002jqsrt..75..765S, 
                                            2002jqsrt..73..557S}.
We followed \citet{2001A&A...376..650Z} 
in assuming an embedding envelope around the core
distribution mimicking the true molecular cloud around \rod.
This outer matter 
shields the core against UV and parts of the optical
outer radiation leading to one magnitude of optical extinction.
As outer radiation field, the mean interstellar radiation field
given in \citet{1994icdi.conf..355B}, an MIR power law distribution
as suggested in \citet{2001A&A...376..650Z}, and the radiation field of
the nearby B2V star were used \citep{1999A&A...344..342L}. 
Numerically, the direction integral was 
performed using $N_d=49$ equally distributed nodes
$(\theta_j,\phi_j)$
on the unit sphere along with optimized integration weights $\omega_j$
\citep{1996jqsrt..56..97S}.

\begin{figure}[t]
\vspace{15.4cm}
\includegraphics{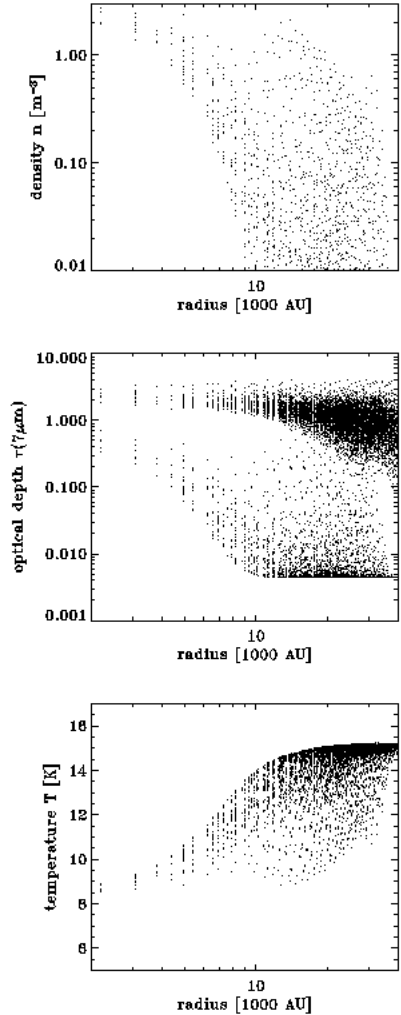}
\caption{Results from a 3D CRT calculation of the density fitting
the MIR maps. The third unknown extent of the single Gaussian was
assumed to be the average of the extent in the two other directions.
The top panel gives the density of the grid cells as a function of
the distance from the southern density peak.
The northern density peaks are visible at a distance of about 20000 AU.
The middle and lower panel show the limits of the
optical depth at 7\ $\mu$m and the
temperature of the grid cells 
as function of the distance from the southern density peak, 
respectively.
\label{Ttau1}
}
\end{figure}
Fig.~\ref{Ttau1} shows results of a first 3D run where we defined for all
Gaussians the HWHM along 
the x-axis (line of sight) to be the mean of the HWHM along the other two
axis. A priori, there is no reason to believe that the distribution along x
is substantially different from the other two axis, so this is an 
appropriate first guess. The Gaussians were not translated along
the line of sight in this run ($t_i=0$).
The top plot in Fig.~\ref{Ttau1} 
gives the density as a function of the distance to the
southern density peak.
As the slope of the structure has different gradients in different
directions, the distribution broadens quickly towards larger radii, with other
maxima at distances of about 20000 AU from the southern peak.
The optical depths for all grid points 
leading from outside the core to the point $\vec x$ along a direction
described by the unit vector $\vec e_{\theta,\phi}$ and the parameterization
$s$ is
\begin{equation}
\tau_{\theta,\phi}(0,\infty,\lambda)=
\sigma(\lambda)\int\limits_0^\infty ds\ 
n\left(\vec x-s\vec e_{\theta,\phi}\right).
\end{equation}
For $\lambda=7\ \mu$m, 
the resulting dependency on the 
distance of $\vec x$ from the southern density maximum is shown in the
middle panel of Fig.~\ref{Ttau1} for all
grid cells. 
Due to the large number of cells and directions, we have plotted, for
each grid cell, only the minimal and maximal optical depth when varying
the direction.
Absorption leads to a maximal optical
depth $<$ 2 for photons reaching or crossing the densest part of the core,
but again the distribution of possible optical depth is broad.
The bottom panel of Fig.~\ref{Ttau1} gives the temperature at all grid
cells as a function of the distance to the density maximum. In the outer
parts, $T$ reaches some 15 K, while at the maximal density, it drops below
9 K. Inbetween, temperatures vary strongly depending on the shielding of
the different grid cells, and no simple approximation of the temperature
can be derived from this plot.

In order to perform a fit of the mm emission of the core density distribution
to the observed mm map, an approximative way to find the temperature 
with moderate numerical effort is required, but not obvious in view
of the large spread in the temperature distributions discussed above.
In the following, we introduce an approximate method to determine temperatures
for pre-stellar cores. Due to the low dust temperatures,
the core emits only in the FIR and mm. As the core is optically
thin to this emission, 
we neglect heating of the inner parts by the outer
low-density parts.
The intensity of the illuminating radiation from the optical to the FIR,
therefore, can be written as
\begin{equation}
I(\lambda,\vec x,\theta,\phi)=
I_{bg}(\lambda)\exp\left[-\tau_{\theta,\phi}(0,\infty,\lambda)\right].
\end{equation}
with the background intensity $I_{bg}(\lambda)$.
The direction integral on the right-hand side of Eq.~(\ref{balance}) can be
interpreted as a weighted averaging over different optical depths, 
so we define a mean optical depth 
\begin{equation} \label{solution}
\overline\tau(\lambda)
\approx
-\ln
\left\{
 \sum_{j=1}^{N_d} \omega_j \exp
 \left[
  -\tau_{\theta_j,\phi_j}(0,\infty,\lambda)
 \right]
\right\}.
\end{equation}
For a point in space being surrounded by matter with equal column
density in all directions, 
the mean optical depth is identical to the optical depth in any
direction
as required. For an embedded point in a core with a gap in a certain
direction $h$, we correctly find
$\overline\tau(\lambda)\approx\tau_{\theta_h,\phi_h}(\lambda)$.
\begin{figure}[t]
\vspace{9.5cm}
\includegraphics{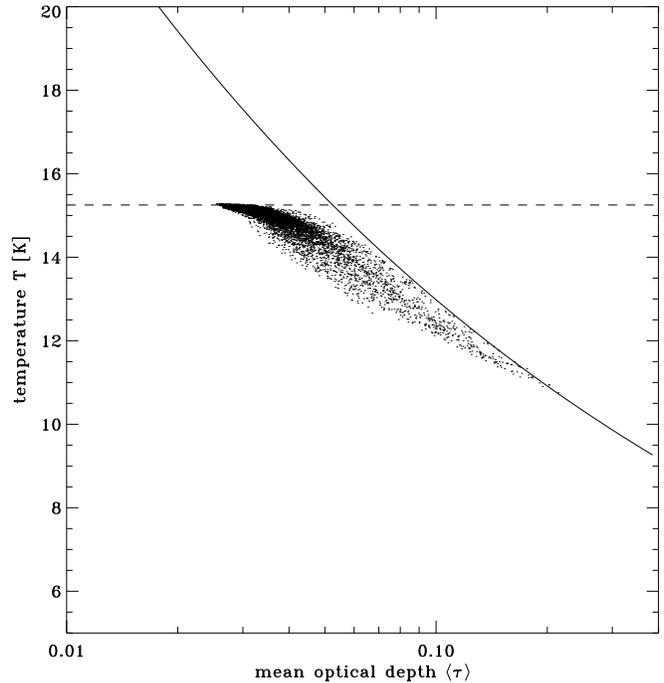}
\caption{Temperature of the grid cells as a function of the mean
optical depth at 7 $\mu$m. The solid line indicates the slope of a 
power law of the form $\tau T^4=$const. The dashed line indicates
the maximal temperature of grid cells outside the core but inside
the cloud.
\label{temtau}
}
\end{figure}

Fig.~\ref{temtau} gives the relation between the temperature and the 
mean optical depth at 7 $\mu$m. Hereinafter, we will call this graph
the "$T_{\overline\tau}$-diagram". The relation shows surprisingly little
spread. For low optical depth $\overline\tau<0.04$, the dust is exposed to the
NIR background radiation field and the derived temperature of 15.5 K
becomes independent of the absorption (dashed line). 
For larger optical depth $\overline\tau>0.1$,
the distribution roughly follows a $\overline\tau\sim T^4$-law (solid line). 
This law depends
on both the background field $I_{bg}(\lambda)$ and the assumed absorption 
coefficient
$\kappa(\lambda)$. In our case, the re-emission of the dust with
a temperature of 7-15 K will occur in the FIR where the assumed absorption
law approximately follows $\kappa\propto\lambda^{-2}$. The integral on the
left-hand side of Eq.~(\ref{balance}) (representing the power density of the
radiation emitted by the dust)
has then an analytical solution 
$\propto \zeta (6)\ \Gamma(6)\ T^6$ with the 
zeta function $\zeta$ and the gamma function $\Gamma$. 
The power density of the incoming radiation
on the right-hand side of Eq.~(\ref{balance}) contains the sum of 
two exponentially-dropping contributions from the NIR and FIR peaks of
the interstellar radiation field, and is 
$\sim \overline\tau_{7\mu \rm m}^{-6/4}$ for the 
background field discussed here. This yields the relation 
$T\sim \overline\tau_{7\mu \rm m}^{1/4}$ as seen in Fig.~\ref{temtau}.
The T($\overline\tau$)-correlation can be found
for other wavelengths as well.
However, for $\lambda<7\ \mu$m, due to the high optical depth, the radiation is
no longer contributing substantially to the heating which broadens the
distribution.
For $\lambda>7\ \mu$m, the relation holds but has more scatter as the
dust is absorbing less efficiently.

We will base the following analysis on assuming that the relation
established in the $T_{\overline\tau}$-diagram will hold even 
when the detailed 3D structure of the
core is changed. The key point is that the heating is mainly related
to the {\em mean optical depth} and not so much to the actual spatial dust
distribution that is leading to this mean optical depth.
\begin{figure}[t]
\vspace{9.5cm}
\includegraphics{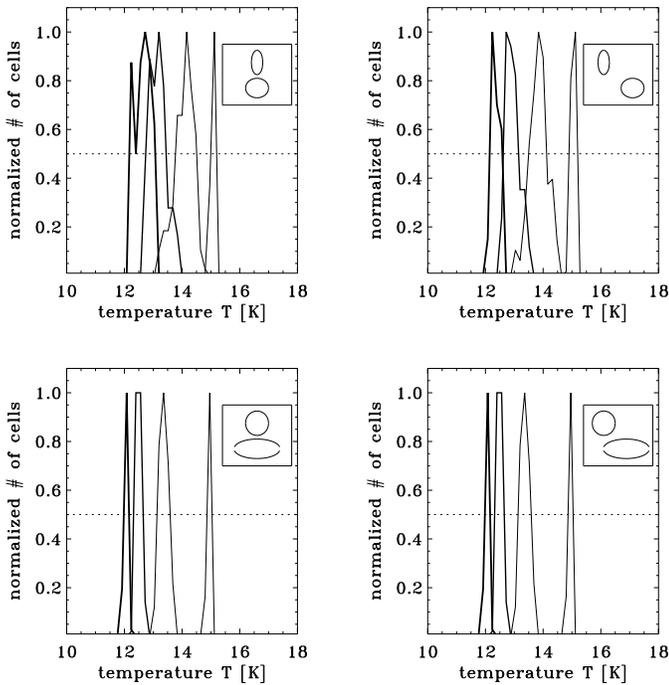}
\caption{Normalized grid cell number
at 7 $\mu$m as a function of the temperature for
an optical depth of 0.03, 0.05, 0.07, and 0.09, respectively, for 4
different configurations sketched in the inset (30 unperturbed Gaussians,
shifted, stretched, and shifted/stretched).
The HWHM can be read from the dashed line. 
\label{HWHM}
}
\end{figure}
We have tested this assumption by moving and deforming
the 30 Gaussian 3D distributions representing our core.
The outcome of the calculations is shown for four typical examples in
Fig.~\ref{HWHM}. To test if the $T_{\overline\tau}$-relation holds
we have plotted the
HWHM of the distribution of grid cells having a temperature T,
corresponding to vertical cuts through the $T_{\overline\tau}$-diagram,
for a mean optical depth at 7 $\mu$m of 0.03, 0.05, 0.07, 0.09, respectively.
For convenience, the peaks were normalized to 1. 
The top left panel gives the HWHM for the non-translated and non-perturbed 
density distribution as derived from the fit of the MIR map and a Gaussian
HWHM along the line of sight
that is the mean of the two other HWHM of the Gaussians.
The right top panel shows the results for a density
distribution where we have 
moved the Gaussians in alternating direction along the line of sight.
In the lower left panel, the Gaussians have been stretched by a factor of
2, and in the lower right panel, they were stretched and shifted 
simultaneously.
Evidently, for all four cases, the HWHM of the grid cell distribution
are around or less than 1 K, with the trend that stretching of the
Gaussians is decreasing the HWHM of the grid cell
distribution. The peaks are moved by less
than 0.6 K compared to the first case.
We conclude that the 
$T_{\overline\tau}$-diagram can be used to determine the temperature
approximately from the mean optical depth for all configurations where the
Gaussians are shifted and stretched within these limiting cases.

\subsection{Fit of the mm map}
To fit the 1.3 mm map of \rod, we applied
the background determination method described already for
the 7 and 15 $\mu$m maps, and obtained a weak
background emission intensity 
$I_{bg}(y,z)$ in the plane of the sky.
The solution of the radiative transfer equation including absorption
and emission along a ray parallel to the x-axis is
\begin{eqnarray} \label{Solution}
&&I(+\infty,y,z)=\cr
&&I_{bg}(-\infty,y,z)\ e^{-\sigma(\lambda)N(y,z)}\ +\cr
&&\sigma(\lambda)\int\limits_{-\infty}^{+\infty}
dx'\ n(x',y,z)\ B_\lambda[T(x',y,z)]\ e^{-\tau_x(x',+\infty,y,z,\lambda)}.
\end{eqnarray}
For wavelengths in the mm range, the optical depth will be small
($\tau_x<10^{-4}$) and the exponentials are approximately 1.
The solution Eq.~(\ref{Solution}) 
contains the unknown number density along the ray $n$ as well
as the temperature distribution $T$. 
Assuming 
3D Gaussians with 30 main axes and 30 relative
translations along the line of sight, we know the full 3D density $n$.
The mean optical depth 
$\overline\tau_{7\mu\rm m}$
can be found and translated into a temperature using the $T_{\overline\tau}$-diagram
by ray-tracing from different directions.
Finally, we integrate along
the line of sight according to Eq.~(\ref{solution}) to find the intensity.
By varying the translation and HWHM of the Gaussians along the line
of sight with the simulated annealing technique, we perform
a fit of the mm map.
\begin{figure*}[t]
\vspace{8cm}
\includegraphics{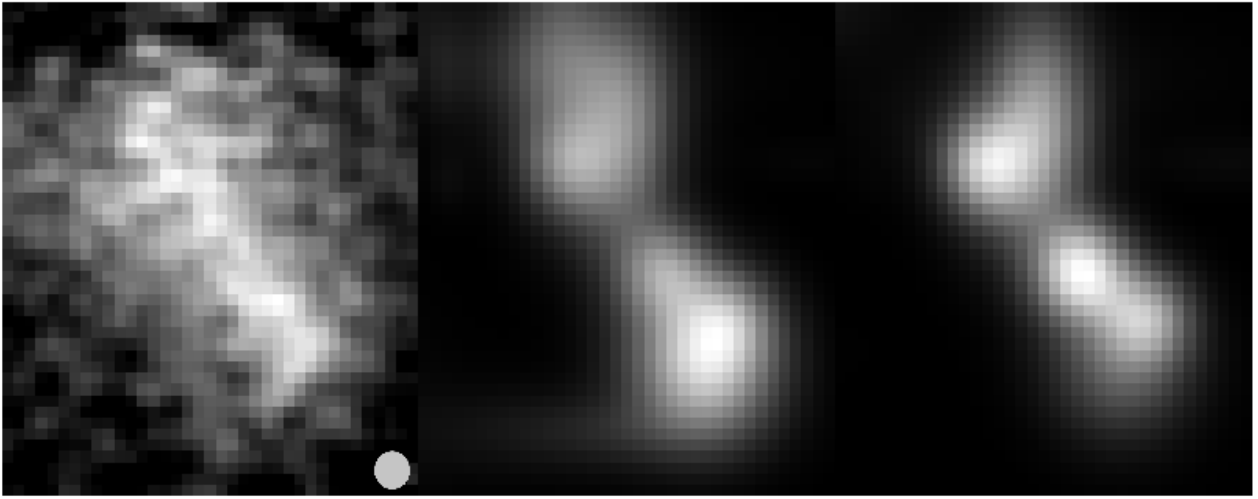}
\caption{Original 1.3 mm map (left), theoretical map for a column density obtained
from the MIR-map fit and a constant temperature of T=10 K (middle), and the 
fitted map using the $T_{\overline\tau}$-relation (right). The grey circle
in the left image represents the beam size.
\label{3maps}
}
\end{figure*}

The resulting fit to the mm map is shown in Fig.~\ref{3maps}. The left
image shows the observed map. Assuming a constant dust temperature of
10 K, the column density map derived from the MIR image can be used to
calculate a corresponding mm map (middle image). Aside from the
noise of the map, comparison of the map reveals several differences.
While the observed map has a more bar-like structure with an approximately
constant flux level along the middle of the bar, the mm map for constant
temperature has a clear gap between the two main emission maxima
seen in MIR. Moreover, the peak flux value of the northern maximum is
lower compared to the maximum flux in the lower part of the image.
Allowing the program to vary the extent of the Gaussians along the
line of sight, the mm emission can be increased by stretching individual
clumps. This way, the optical depth to the considered point in space is
lowered allowing a more efficient heating and a rise in temperature.
While the density is slightly lowered by the stretching, the overall 
effect is to increase the emissivity. 
The optimization program therefore stretched the Gaussians
in regions where the observed image was brighter, especially
in-between the two MIR-emission maxima. Also, the Gaussians 
forming the maximum in the northern part of the image have been stretched
and moved away from each other to decrease the mean optical depth and to
increase the emissivity.
The right image in Fig.~\ref{3maps} shows the results of shifting
and stretching/compressing the Gaussians along the line of sight
to fit the mm map.
The emission pattern now is more bar-like, but still a gap
is visible that cannot be seen in the observed picture. The brightness
of the two maxima is equalized, and flux is moved from
the southern emission maximum towards the middle structure.

Discussing the deviations, one important point has to be raised
in the beginning. A much better fit could be achieved when
the Gaussians would be allowed to move in the plane of sky. But as
the method is designed to fit both the MIR and mm maps,
the Gaussians are fixed in the plane of the sky to enforce a
simultaneous fit of the MIR map. This includes that the total
mass of each Gaussian
is fixed. 

The freedom in the fit is reduced due to the limited
number of Gaussians, and a better fitting especially in-between the
two MIR-emission maxima could be expected from a run with more
Gaussians. 3D radiative transfer, even when using the $T_{\overline\tau}$-method
and combined use of simulated annealing is at the limit of the
currently available computer powers so that a substantial increase of the
number of Gaussians is not feasible.
In addition, the low-resolution mm map used for the fit has a sub-structure that 
prohibits unambiguous identification of fine-structure in the
image. We hope to improve this
with upcoming new mm maps.
A deeper discussion of the various uncertainties, errors, and
approximations is given in the next section.

\begin{figure}[t]
\vspace{15cm}
\includegraphics{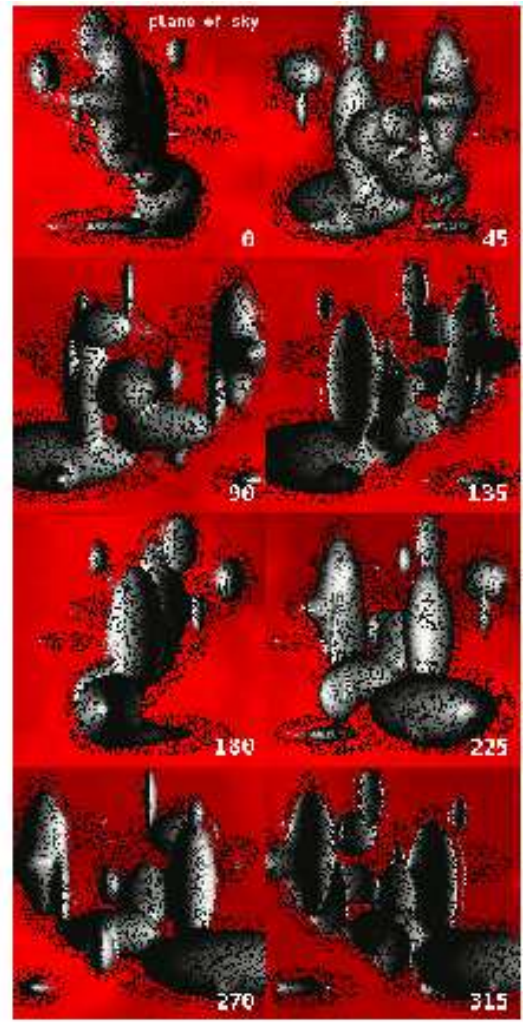}
\caption{Iso-density surfaces of the 3D dust density structure
of the cloud core \rod\ 
at 5\% (solid) and
1 \% (semi-transparent) of the maximal density for different
viewing angles ranging from 0 to 320 degrees with respect to the line
of sight seen against an artificial background. 
\label{rotdensity}
}
\end{figure}
In Fig.~\ref{rotdensity}, we have visualized the 3D dust density structure
of the core by showing two iso-density surfaces at 5\%
(solid surface) and
1 \% (semi-transparent surface) of the maximal density
and rotating it against an
artifical background.
The lower density maximum is compact with an extension to
model the MIR emission seen in the left lower part of the MIR image.
The northern structure is more complex and extended to model both the
broader structure seen in the MIR and the high emissivity seen in the
mm. 
In Tab.\ref{coreprop}, we summarize the overall properties
of the cloud.
   \begin{table}
      \caption[]{Derived overall properties of the core \rod}
         \label{coreprop}
     $$
         \begin{array}{p{0.5\linewidth}l}
            \hline
            \noalign{\smallskip}
            Spatial extent   & 0.22\ {\rm pc}\\
            Total gas mass   & 2.3\ {\rm M}_{\sun}\\
            Total gas mass of northern peak & 0.8\ {\rm M}_{\sun}\\
            Total gas mass of southern peak & 1.5\ {\rm M}_{\sun}\\
            Maximal density at southern peak & 30\ {\rm m}^{-3}\\
            Main axes aspect ratio               & 2:2:1\\
            Mean gas density & 8\times 10^{-16}\ {\rm kg\ m}^{-3}\\
            Temperature range & 10-15\ {\rm K}\\
            \noalign{\smallskip}
            \hline
         \end{array}
     $$
   \end{table}
%

\begin{figure}[t]
\vspace{10cm}
\includegraphics{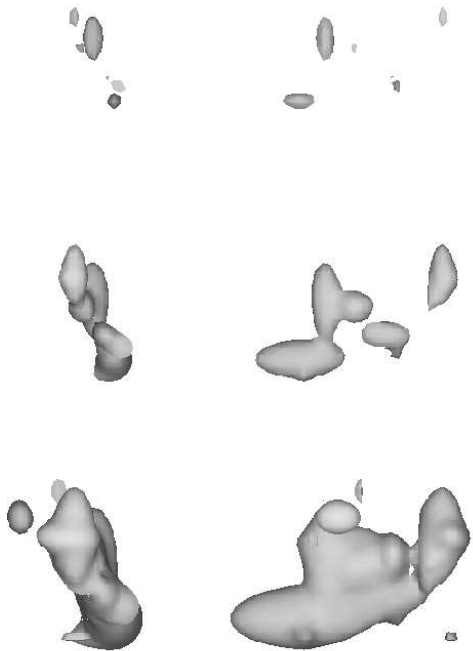}
\caption{Iso-temperature surfaces of the 3D dust temperature structure
of the cloud core \rod. The left panel shows the distribution seen
along the line of sight, while, for the right panel, the view has
been rotated by 90$^\circ$. The iso-temperatures are 12 K,
12.8 K, and 13.8 K, respectively (top to bottom).
\label{rottemp}
}
\end{figure}
The 3D dust temperature distribution is visualized in Fig.~\ref{rottemp}
by showing iso-temperature surfaces for increasing temperature.
The left panel gives the view along the line of sight (corresponding
to the top left image in Fig.~\ref{rotdensity}), while in the
right panel, the configuration is rotated by 90$^\circ$ in a plane 
perpendicular to the plane of the sky (corresponding
to the image with the label "135" in Fig.~\ref{rotdensity}). 
The iso-temperatures are 12 K,
12.8 K, and 13.8 K, respectively (top to bottom).
In general, the distribution follows the density distribution, as the
densest parts are well-shielded and have low temperatures. 
In the northern part, however, there is a low-temperature region
that is not only caused by shielding due to matter around the local
density maximum, but also by other nearby local density maxima.
This effect becomes important when the density distribution consists
of many clumps with comparable maximal density values as in the
northern clump.
We note that such a realistic temperature distribution including
shielding effects can only be obtained doing RT calculations, but
that the proposed fast $T_{\overline\tau}$-method includes this effect.
Animations showing the density and temperature structure
are displayed at 
http://www.mpia-hd.mpg.de/homes/stein/Ani/movief.htm.
%
%
%
%
\section{Ambiguity and error discussion for the 3D modeling}
Deriving the complex 3D structure of pre-stellar cores from
projected images is a difficult and challenging task.
The method and results presented in this paper are a first step
to enter this rich 3D world.
While the advantage over a 1D or 2D approach is obvious in view of
the complex structures of observed cloud cores, 
the various errors 
arising from approximations and limited knowledge of the
physical conditions make a careful
error analysis inevitable. Lacking any analytical solution, however,
a quantitative error estimate is often not possible.
In turn, many of these errors or ambiguities might be reduced
when maps of higher resolution, sensitivity, and dynamical range
at the considered wavelengths or images at other
wavelengths become available (e.g. by SPITZER, HERSCHEL, JWST, ALMA), or
when independent information, e.g., about the background radiation
field can be added to the currently available information.

In the following, we will discuss the ambiguities and errors arising
from usage of
the numerical algorithms, from the approximations and assumptions, 
and from the observational errors. 

\subsection{Numerical errors}
The column density and emission integrations for given density and
temperature distributions are performed with a 5th-order Runge-Kutta scheme
incorporating adaptive step size control. The error of this ray-tracing
can be neglected compared to the other sources of error.

The density distribution contains 30 three-dimensional Gaussians.
In Sect.~6, we already concluded that the number of Gaussians might
be too low to resolve the material between the two MIR peaks
sufficiently, causing deviations in the mm emission from this region
of the core. We did extensive tests including Gaussians
with arbitrary orientation in space. Indeed, a smaller number of
Gaussians of this type is required to fit a density distribution than
for Gaussians with axes along the Cartesian axes. However, the formula
for Gaussians with arbitrary orientation in space is substantially more complex
increasing the run time of the fitting routine beyond current computer
capabilities. We are currently elaborating, if the code can be accelerated
by other means to be able to increase the number of Gaussians.

The 3D CRT code has been tested in the framework of the 2D spectral
benchmark published in \citet{2004A&A...417..793P}, the first benchmark
to test five Monte-Carlo and grid/ray-tracing CRT codes by calculating
the spectral energy distribution of a star surrounded by a circumstellar
dust disk. The deviations of the results
for optical depths considered here were below
10\%, and most of the deviations were caused by the scattering term, whereas
scattering plays no important role for the heating of pre-stellar cores.
Nevertheless, we point out that there is no 2D CRT benchmark comparing
images nor a 3D CRT benchmark of any kind.

The limitations and possible errors of the proposed $T_{\overline\tau}$-method
have been discussed already in Sect.~4 for the application to the cloud
core \rod\ with an approximate total error of about 1 K or less. 
In general, the method requires to rerun the full 3D CRT
code for any newly considered background radiation field
in order to obtain the $T_{\overline\tau}$-relation. 
In cases of strong sub-structure of the core material, the number of considered
directions for the averaged optical depth might have to be increased. The possible
errors arising from additional, but unresolved shielding can be neglected
in most cases, but for structures with holes, the method might underestimate
the temperature when missing to resolve the illumination through the hole.
Comparing the scale of the sub-clumps with the spatial resolution of 1000 AU/pixel for
cores in $\rho$ Oph with MIR-instruments like ISOCAM, it is evident, though,
that such a strong sub-clumping is currently not observed.

Operating in a complex 210-parameter space, the main question is the
ambiguity of the derived optimized parameter set. Optimization
algorithms like the gradient search usually succeed in 
finding a local minimum, but often
fail in localizing the deepest minimum of the $\chi^2$-function of the fit.
For this purpose, more sophisticated methods like simulated annealing or
genetic algorithms are applied which have a higher probability in finding
the deepest minimum. A prove for completeness is impossible in our case,
but from our former experience in applying the method \citep{1994A&A...287..493T,
1996jqsrt..56..97S,2003ApJ...583L..35S} as well as from exploring different
initial parameter sets, we are confident that the solutions are close to the density 
structure presented here. Nevertheless, it is correct to state that we can
only show that the derived density structure simultaneously fits the three
images discussed here, but that we
can not exclude that there is another distribution
with a comparable $\chi^2$-value. We note that the presented density distribution
is a valuable working hypothesis for further studies and that we intend to remove
some of the ambiguity when more data become available.

The proposed method has another ambiguity that is common in interpreting
projected images of 3D configurations. As soon as Gaussians are
separated along the line of sight (e.g. when the distance of the maxima is
much larger than the HWHM) and not indirectly connected by other Gaussian, 
there is no way to determine the distance of
the clumps.
This is because the mm fit uses the temperature difference arising from
overlapping Gaussians, and therefore, the temperature is constant for all clumps
of large separation.
Moreover, separated clumps can be exchanged in their position along the
line of sight without changing the observed MIR and mm flux.

For the density structure derived here, this is the case only
for a few low-density
Gaussians. Hence, we conclude that this ambiguity does not affect the
derived distribution of \rod.
%
\subsection{Approximation and assumption errors}
The assumptions concerning dust properties have already been discussed
in Sect.~3 and the beginning of Sect.~4. 
As the total dust mass derived from the MIR maps does not strongly depend
on the particle size, we expect the uncertainties in the optical properties
of the dust and the gas/dust ratio to dominate the error in the mass.
It should be noted that another small error has been introduced by
comparing a flux that has been observed through the wavelength filters
of ISOCAM to images calculated for a single wavelength.
An advanced modeling of the dust properties will be subject
to a forthcoming project.

It might be objected that the expansion of the density distribution
may cause substantial errors when fitting filamentary structures,
especially when inclined to the Cartesian axes.
Firstly, the images with the 1000 AU/pixel resolution do not resolve
such strongly elongated structures.
Secondly, in order to avoid un-physically flat structures,
the fitting algorithm allows main axes ratios of up to 10.
The maximal ratio obtained
in the fits is about 5. Moreover, elongated structures inclined to
the Cartesian axes are fitted by several Gaussians with moderate
main axes ratios. We therefore argue that the expansion into 3D Gaussians
does not produce large fitting errors.

For the determination of the column density from the MIR images, we
assume that the invisible background behind the core can be
extrapolated from the visible background in the vicinity of the core
(the details of the used 2 step-interpolation are given in Sect.~3).
The stability of the method was tested by varying interpolation radius
and core area. 
The local variations within the interpolation radius
are of the order of 10-15\% arising either from variations in the
illuminating MIR flux of the photo-dissociation region or from low-density
molecular cloud structures along the line of sight.

The exact temperature distribution depends on the assumed properties of
the external illuminating radiation field. The 3D radiation
field for a core within $\rho$ Oph is not known precisely, as the
position within the cloud along the line of sight is unknown, and
the MIR flux of the photo-dissociation region and nearby other
cores has not been modeled in detail. We note that our approach to
have a composition of interstellar radiation field, MIR power law
component and nearby B2V star is more realistic than a 1D background
field as it is assumed in 1D fits.
\subsection{Observational errors}
The noise of the MIR images is about 0.3 MJy/sr for the LW2
ISOCAM filter, so that the MIR flux is uncertain to a relative error
of less than 10 \%. 
For the mm image, the noise is 6 mJy/13" beam for the core center
and the relative flux error is of the order of 15 \%.

The simultaneous fit of different images requires an accurate
pointing. This is especially valid for the comparison
of the MIR and mm image, as the Gaussians are not free to move
in the plane of sky when performing the fit to the mm map.
An offset would cause the fitting procedure to compensate 
deviations in position by varying the temperature un-physically.
The absolute pointing error in the images was less than 1.4" (2$\sigma$)
for the ISO telescope and 3" (1$\sigma$) for the IRAM 30m
telescope,
which is less than the typically obtained HWHM of the
Gaussians.

The technique used for the IRAM 30m 1.3 mm bolometer observations
("dual beam") is known to filter out some of the extended emissions
\citep[e.g.][]{1996A&A...314..625A}.
\citet{2001A&A...365..440M} have simulated the effects of the 
dual-beam technique on model objects with intensity profiles following
simple power laws and calculated the loss of flux arising in these
observations. The flux loss is greater for low power-law indices and
increases with increasing radius. We used their modeling to evaluate
the flux loss in \rod\ and found it to be negligible.

In Sect.~3, we have used a zodiacal light component being constant over
the observed area which has an uncertainty of about 20\%. Due to
the large contribution of the zodiacal light to the total flux and
its large error, this foreground flux is dominating the uncertainties
in the modeling that arise from observational errors.
\section{Comparison with 3D density distributions from gravo-turbulent
star-formation models}
In the picture of gravo-turbulent star formation, 
molecular cloud cores are not quasi-static equilibrium
objects. Rather they are dynamically evolving as part of the overall turbulent
cascade. The same complex and stochastic flow that forms a core at first place
continuously reshapes its structure and even may disperse it again.
Those cores with an excess of gravitational energy collapse rapidly to
form stars, while the others with sufficiently large internal or
kinetic energies re-expand once the turbulent compression subsides
\citep{1996A&A...313..269T, Vas04}.
Consequently, molecular cloud cores are observed having a variety of different
shapes, ranging from highly elongated filaments via cometary-shaped structures
to very roundish spherically symmetric cores \citep[see e.g.][]{1999pcim.conf..227M}. 
This sequence
roughly reflects the increasing importance of self-gravity compared to
turbulent kinetic energy. Objects that are dominated by turbulent ram
pressure tend to have on average more complex morphological appearance than
more quiescent cores on the verge of gravitational collapse \citep{Klessen04}.
The latter usually have a rather generic density structure with flat
inner profile, followed by an approximate power-law fall-off further out, and
apparent truncation at some maximum radius.

\begin{figure*}[t]
\vspace{6cm}
\includegraphics{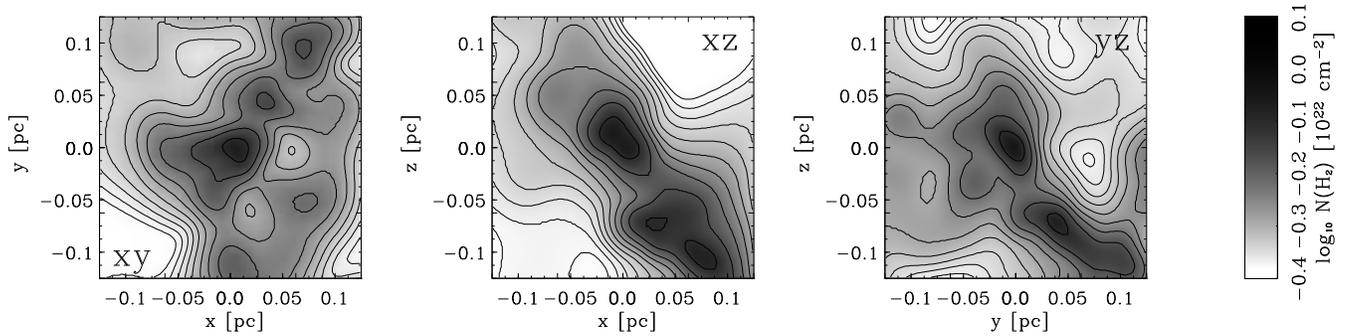}
\caption{Column density map of a typical pre-stellar core based
on a SPH simulation of a self-gravitating supersonically turbulent molecular
cloud. The density is integrated along the three Cartesian axes.
\label{SPH}
}
\end{figure*}

The double-peaked and elongated nature of the core \rod\ places it
somewhere in between these two extremes. Its 3D density structure is typical
for pre-stellar cores in their early phases of evolution when the dynamics is
still dominated by external compression rather than by self-gravity \citep{Klessen04}.
The relatively small density contrast of about 30 is
additional evidence for that.  To illustrate this point, Fig.~\ref{SPH}
shows the column density map of a typical pre-stellar core in a
smooth particle hydrodynamics (SPH) simulation 
of a self-gravitating supersonically turbulent molecular
cloud for comparison (integration along the Cartesian axes).
  
The computation focuses on a cubic volume within the cloud and follows the
evolution using smooth particle hydrodynamics 
\citep[SPH; see][]{1990nmns.work..269B, 1992ARA&A..30..543M}.
This is a particle-based scheme to solve the equations of hydrodynamics
well-suited for complex flows with large density contrasts. Turbulence is
continuously driven on large scales using a non-local Gaussian field such that
the root mean square Mach number is constant \citep{1999ApJ...524..169M} 
with value ${\cal M} \approx
6$.  Details of the numerical method and associated resolution issues are
discussed, e.g., by \citet{2000ApJS..128..287K} and
\citet{2000ApJ...535..887K}. The
gas is taken to be isothermal with temperature $T=11.3\,$K, corresponding to a
sound speed $c_{\rm s} = 0.2\,$km$\,$s$^{-1}$. 
The computational domain corresponds to a cube of
$2.5$ pc length, containing  $1140\ M_{\odot}$ of
molecular cloud gas. Note, Figure 12 shows
only a fraction of $1/103$ of the total volume  centered on
the considered protostellar core.
Complementary aspects of the adopted model are discussed by
\citet[their model LSD at time $t_1$,][]{2003ApJ...592..188B},
\citet[model M6k2,][]{2004A&A...419..405S},
\citet[model M6k2,][]{2004A&A...423....1J}, and
\citet[model LSD,][]{Klessen04}.

The model core in Fig.~\ref{SPH} is selected because its morphological
appearance closely resembles that of \rod. At the depicted time
it is gravitationally unstable and will collapse to build up a protostar
during its subsequent evolution. However, due to the highly stochastic nature
of compressible turbulent flows the fate of a molecular cloud core is
predictable only in a statistical sense.  The selected model core, therefore,
can at best be indicative of what may happen to \rod\ in the next
$10^5$ years.  Nevertheless, it is fair to speculate that gas around 
the southern density maximum will go into collapse. The fate of the
more extended and less dense northern core is hard to predict without
further study of possible infall motion. When collapsing into two fragments, 
the core \rod\ would give rise to a wide binary system. A
detailed investigation of the dynamical evolution of the core based
on the observational data will be the subject of future studies.
This will require a combined modeling of the continuum and line
data to incorporate velocity informations.
\section{Conclusion and Outlook}
In this paper, we propose a new method to model the three-dimensional
dust density and temperature structure
of dense molecular cloud cores.
The method is based on the fits of multiple continuum images in which
the core is seen in absorption and emission
and requires only a few computations with a 3D CRT
code. The large parameter space of a three-dimensional 
structure is covered using simulated annealing as optimization algorithm.
One of the key points of the proposed method is the use of a $T_{\overline\tau}$-relation
that has been derived from the 3D CRT runs. It links
the mean optical depth at a given wavelength from outside to a spatial
point within the core to the dust temperature at this point, 
making a fast evaluation of the mm emission along the
line of sight possible.

We have applied the method to model the dense molecular cloud core
\rod. In the MIR, the core is seen in absorption against a bright
background from the photo-dissociation region of the nearby B2V star,
illuminating the cloud from behind.
Two ISOCAM images 
at 7 and 15 $\mu$m have been fitted simultaneously
by representing the dust
distribution in the core with a series of 3D Gaussian
density profiles. The background emission behind the core was
interpolated from nearby regions of low extinction.
Using simulated annealing, we have obtained a 2D column density
map of the core.
The column density of the core
has a complex elongated pattern with two peaks,
with the southern peak being more compact.
To retrieve the full three-dimensional structural information,
we have calculated the temperature structure of the core with a 3D CRT
code, assuming
several limiting cases of core extent along the line of sight.
It was shown that, for a given position in the core, 
the relation between the dust temperature and the mean
optical depth from outside to this position varies within less than 1 K,
when changing the shape of the core along the line of sight.
Therefore, the $T_{\overline\tau}$-relation was used to estimate quickly
the temperature at a given point by a fast direction averaging
over the optical depth at just one wavelength.
Using this mean, we have varied extent and position
of the Gaussian components of the density along the line of sight
until a fit of both the MIR and a 1.3mm map, obtained with the IRAM 30m telescope,
could be achieved.
We have presented the three-dimensional dust density and temperature
structure, revealing a condensed southern part and a more extended
and complex northern part.
We have addressed several sources of errors, namely the background
determination method, the assumed dust particle properties, errors
in the maps, the ambiguity in the derived distribution, and the
approximation error of the $T_{\overline\tau}$ method.
As we did not reach a perfect fit of the mm image between the two
maxima due to the low number of Gaussians, we expect that the detailed
3D structure in-between the density maxima might change when using a
higher number of Gaussians. The general structure of a condensed southern
maximum and a complex northern multi-clump region, however, is clearly
resolved within the limits of the presented analysis.
This structure is in general agreement
with recent gravo-turbulent
collapse calculations for molecular clouds.
We speculate that the southern condensation belongs
to the category of cloud
clumps which are dominated by self-gravity. It may be collapsing to form a star in
the near future.
The northern clumps, however,  may still have sufficiently large internal or
kinetic energy to re-expand and merge into the ambient medium 
once the turbulent compression subsides.

In this application, we determine about 210 free parameter to fit about 2000 pixels
simultaneously, but hidden in the assumptions are many more free parameter
like the dust properties or details of the illuminating background.
It has to be kept in mind, though, that the hidden parameters are present
in any other, more simple modeling as well, and often with more unrealistic
approximations. The $\rho$ Oph cloud is illuminated from behind in contradiction
to a 1D boundary condition for the incoming
radiation.

The perspective of applying the method is promising. There are a number
of well-observed cores where images at more than two wavelengths are
available. With every additional image, the ratio of
new unknown background parameter versus constrained parameter decreases, as
the density parameters are unchanged. This will increase the accuracy
of the determined density structure. Moreover, the derived $T_{\overline\tau}$-relations
for \rod\ can be used in other applications incorporating the temperature.

The ultimate goal of applying the method to well-observed cores, however,
will be to address the key question of early star formation, namely if the
considered cores have in-falling material. 
The current line observations provide the molecular line emission flux 
integrated over all moving gas cells along the line of sight. 
In the general case, gas motion and emissivity of the cells can not be
disentangled, and the 1D approximation or shearing layers are assumed
to unfold it. Without unfolding it, infall motion can be mixed up with
rotational motion leaving it undecided if the core shows any sign for the 
onset of star formation.
This is changed if the core has been investigated with the $T_{\overline\tau}$-method.
Knowing the full 3D structure in dust density and temperature, the line
of sight-integral can be inverted providing the complete kinematical 
information if the considered line is optically thin and a model
for the depletion of the considered molecules is used.
This direct verification of infall motion will be subject to a forthcoming
publication.

\acknowledgements 
We gratefully acknowledge stimulating discussions with Ralf Launhardt,
Tigran Khanzadyan, Sebastian Wolf, and Ulrich Klaas. R.S.K. acknowledges
support from the Emmy Noether Program of the Deutsche 
Forschungsgemeinschaft (DFG: grant no.\ KL1358/1).
This research has made use of NASA's Astrophysics Data System Abstract Service.

\bibliographystyle{aa}
\bibliography{rhoophd}

\end{document}